\documentstyle[12pt,intlim]{amsart}
\def\rtimes{\mathbin{\times \mkern-5mu{\footnotesize |}}}
 \theoremstyle{definition}
  
\newcommand{\cO}{{\cal O}}

\begin{document}

\size {12}{18pt}\selectfont

\title[The Universal Vassiliev-Kontsevich invariant]{\bf
The universal Vassiliev-Kontsevich Invariant for framed oriented links}

\author[T. Q. T. Le \quad\& \quad J. Murakami] { Le Tu
Quoc Thang and Jun Murakami} \address{Max-Planck Institut f\"ur
Mathematik, Gottfried-Clarenstr., 26, 53225 Bonn, Germany}
 \email{letu@@mpim-bonn.mpg.de\\ jun@@math.sci.osaka-u.ac.jp}
 \date{December 1993}
\maketitle \begin{abstract} We give a generalization of the
Reshetikhin-Turaev functor for
 tangles to get a combinatorial formula for the  Kontsevich integral
 for framed  oriented links. The uniqueness of the universal
 Vassiliev-Kontsevich invariant  of framed oriented
 links  is established. As a  corollary one gets the
 rationality of Kontsevich integral.\end{abstract} \parskip = 4pt plus
1pt \newcommand{\Zf}{\mbox{$Z_f$}} \centerline{ INTRODUCTION}

This is an expository paper on the construction of the universal
Vassiliev-Kontsevich invariant of framed oriented links. We give a
description of a generalization of the  Reshetikhin-Turaev functor.
This is a mapping from the set of all framed oriented tangles to
rather complicated sets. This mapping, when restricted to the set of
all framed oriented links in  3-sphere $S^3$, is an isotopy invariant
called the universal Vassiliev-Kontsevich invariant of framed oriented
links.  It is as powerful as the set of all invariants of finite type
(or Vassiliev invariants) of framed oriented links. Hence it dominates
all the invariants coming from quantum groups in which the $R$-matrix
is a deformation of identity, as in \cite{Re-Tu,Tu}. Similar
constructions of the universal Vassiliev-Kontsevich invariants appear
in \cite{Car,Piu}.

The values of the universal Vassiliev-Kontsevich invariant of framed
knots lie in an algebra, and if we project to an appropriate quotient
algebra, we get the Kontsevich integral of knots (Theorem 6).

Actually the universal Vassiliev-Kontsevich invariant is constructed
using an object called the  Drinfeld associator. This is a solution of
a system of equations. Every solution of this system gives rise to a
universal Vassiliev-Kontsevich invariant which we prove (Theorem 8)
is independent of the solusion used. As a corollary we get the
 rationality of the
universal Vassiliev-Kontsevich invariant and Kontsevich integral.

The rationality of Kontsevich integral was claimed in \cite{Kont},
without proof, citing only Drinfeld's paper \cite{Drin2}. The result of
\cite{Drin2} can not be applied directly to this case because the
spaces involved, though related, are in fact different. Here we modify
Drinfeld's proof to our situation, using a suggestion of Kontsevich.

Many properties of the universal Vassiliev-Kontsevich invariant are
 established.  Connections to quantum group invariants and to
multiple zeta values are discussed.

For a detailed exposition of the theory of the Kontsevich integral
 and the
universal Vassiliev-Kontsevich invariant for (unframed) knots see
\cite{Bar}. Many arguments in \cite{Bar} are generalized here.

For technical convenience we use $q$-tangles instead of tangles. This
concept is similar to that of a $c$-graph  introduced in \cite{Coste}.
Actually the category of $q$-tangles and the category of tangles are
the same, by Maclane's coherence theorem.

\noindent{\bf Acknowledgement} We are sincerely grateful to M.
Kontsevich for many useful discussions.  We would like to thank H. Boden
for careful reading the manuscript and correcting many mistakes. The
authors thank the Max-Planck Institut f\"ur mathematik for hospitality
and support.

\section{Chord diagrams} Suppose $X$ is a one-dimensional compact
oriented  smooth manifold whose components are numbered. {\it A chord
diagram with support $X$} is a set consisting of  a finite number of
 unordered pairs
of  distinct non-boundary   points on $X$, regarded up to orientation
 and component preserving homeomorphisms.  We view each
pair of points as a chord on $X$ and represent  it as a dashed line
connecting the two points. The points are called the vertices of
chords.  \newcommand{\id}{\text{id}} \newcommand{\Pf}{{\bf Proof: }}
\newcommand{\cA}{{\cal A}} \newcommand{\cX}{{\cal X}}
\newcommand{\cC}{{\cal C}} \newcommand{\cG}{{\cal G}}
\newcommand{\cD}{{\cal D}}
\newcommand{\g}{{\frak g}}
\newcommand{\R}{{\Bbb R}} \newcommand{\Q}{{\Bbb Q}}
\newcommand{\N}{{\Bbb N}} \newcommand{\Z}{{\Bbb Z}}
\newcommand{\ve}{{\varepsilon}} \newcommand{\C}{{\Bbb C}}
\newcommand{\Pro}[1]{\noindent {\bf Proposition #1}}
\newcommand{\Thm}[1]{\noindent {\bf Theorem #1}}

Let $\cA(X)$  be the vector space over $\Bbb Q$ (rational numbers)
spanned by all chord diagrams with support $X$, subject to the 4-term
relation:  $$ \raisebox{-15pt}{ \begin{picture}(63,50)(-7,0)
\thicklines \put(15,0){\vector(1,0){15}}
\put(9,37.5){\vector(-3,-4){9}} \put(45,25.5){\vector(-3,4){9}}
\end{picture}}-\raisebox{-15pt}{ \begin{picture}(63,50)(-7,0)
\thicklines \put(15,0){\vector(1,0){15}}
\put(9,37.5){\vector(-3,-4){9}} \put(45,25.5){\vector(-3,4){9}}
\end{picture}}+\raisebox{-15pt}{ \begin{picture}(63,50)(-7,0)
\thicklines \put(15,0){\vector(1,0){15}}
\put(9,37.5){\vector(-3,-4){9}} \put(45,25.5){\vector(-3,4){9}}
\end{picture}}-\raisebox{-15pt}{ \begin{picture}(63,50)(-7,0)
\thicklines \put(15,0){\vector(1,0){15}}
\put(9,37.5){\vector(-3,-4){9}} \put(45,25.5){\vector(-3,4){9}}
\end{picture}} =0$$

$\cA(X)$ is graded by the number of chords. We denote the
completion with respect to this grading also by $\cA(X)$.

Every homeomorphism $f:X\rightarrow Y$  induces an isomorphism between
$\cA(X)$ and $\cA(Y)$.

\newcommand{\cB}{{\cal B}} On the plane $\R^2$ with coordinates $(x,t)$
consider the set $X$  consisting of  $n$ lines $x=1$, $x=2$,$\ldots$,
$x=n$, lying between two horizontal lines $t=0$ and $t=1$.  All the lines
 are oriented
downwards. The space
$\cA(X)$ will be denoted by $\cB_n$. A component of $X$ is called a string.
 The vector space
$\cB_n$ is an algebra with the following multiplication. If $D_1$ and
$D_2$ are two chord diagrams in $\cB_n$ then $D_1\times D_2$ is the
chord diagram gotten by putting $D_1$ above $D_2$. The unit is the
chord diagram without any chord. Let $\cB_0=\Bbb Q$.

\Pro{1:}\cite{Kont}{\it The algebra $\cB_1$ is commutative.}

When $S^1$ is a circle, $\cA(S^1)$ is  denoted  simply by $\cA$.

Suppose $X,X'$ have  distinguished  components $\ell,\ell'$, $X$
 consists of
 loop components only. Let $D\in\cA(X)$ and $D'\in\cA(X')$. From each
 of $\ell,\ell'$ we remove a small arc  which does not contain any
 vertices.
The remaining part of $\ell$ is an arc which we glue  to $\ell'$ in
the place of the removed arc such that the orientations are compatible.
The new chord diagram is called {\it the connected sum of $D,D'$
along the
distinguished components}. It does not depend on the
locations of the removed arcs, which follows from  the 4-term
relation and
the fact that all components of $X$ are loops. The proof is
the same as in case $X=X'=S^1$ as in \cite{Bar}.

In case when $X=X'=S^1$, the connected sum defines a multiplication which
turns  $\cA$ into an algebra. This algebra is isomorphic to
 $\cB_1$ (cf. \cite{Bar,Kont}).

Suppose  again $X$ has a distinguished component $\ell$. Let $X'$ be
the manifold gotten from $X$ by reversing the orientation of $\ell$. We
define a linear mapping $S_\ell:\cA(X)\rightarrow\cA(X')$ as follow. If
$D\in\cA(X)$ represents by a diagram with $n$ vertices of chords on
$\ell$. Reversing the orientation of $\ell$, then multiplying by
$(-1)^n$, from $D$ we get the chord diagram $S_\ell(D)\in\cA(X')$. Note
that $S_\ell:\cB_1\rightarrow\cB_1$ is an anti-automorphism.

Now let us define $\Delta_i:\cB_n\rightarrow\cB_{n+1}$, for $1\le i\leq
n$.  Suppose $D$ is a chord diagram in $\cB_n$ with $m$ vertices on the
$i$-th string.
Replace the $i$-th string by two strings, the left and  the right, very
close to the old one. Mark the points on this new set of $n+1$ strings
just as in $D$; if a point of $D$ is on the $i$-th string then it
yields two possibilities, marking on the left or on the right string.
Summing up all possible chord diagrams of this type, we get
$\Delta_i(D)\in\cB_{n+1}$.

Define $\ve_i$ by $\varepsilon_i(D)=0$ if the diagram $D$ has a
vertex of
chords on the $i$-th string. Otherwise let $\varepsilon_i(D)$ be the
diagram in $\cB_{n-1}$ gotten by throwing away the $i$-th string.
 We continue
$\varepsilon_i$  to a  linear mapping from $\cB_n$ to $\cB_{n-1}$.

Notation: we will write $\Delta$ for $\Delta_1:\cB_1\rightarrow \cB_2$,
 $\id\otimes\ldots\otimes\Delta\otimes\ldots\otimes \id$ (the $\Delta$
 is at the $i$-th position) for $\Delta_i$;  $\varepsilon$ for
 $\varepsilon_1:\cB_1\rightarrow \cB_0=\Bbb Q$,
 $\id\otimes\ldots\otimes\varepsilon\otimes\ldots\otimes \id$ (the
 $\varepsilon$ is at the $i$-th position) for $\varepsilon_i$.

\noindent{\bf Remark:} The reader should not confuse $\Delta$ with
the co-multiplication introduced in \cite{Bar} for $\cA$.

\Pro{2:} {\it  We have \begin{equation} (\Delta\otimes
\id)\Delta=(\id\otimes\Delta)\Delta\end{equation}} This follows easily
from the definitions.

Put $\Delta^n=\underbrace{(\Delta\otimes \id\otimes\ldots\otimes
\id)}_{\mbox{$n$ times}}\underbrace{(\Delta\otimes \id\ldots\otimes
\id)}_ {\mbox{$n-1$ times}}\ldots(\Delta\otimes \id)\Delta$. For $n=0$
 let $\Delta^n=id:\cB_1\to\cB_1$.

\Thm{1:} {\it The image of $\Delta^n:\cB_1\rightarrow\cB_{n+1}$ lies
in the
 center of $\cB_{n+1}$.}

The proof is not difficult, it can be proved by imitating the case
 $n=0$ which
is Proposition 1 and is proved in \cite{Bar}.

If $D\in\cB_n$ then $1^{\otimes n_1}\otimes D\otimes 1^{\otimes n_2}$
is the element of $\cB_{n_1+n+n_2}$ which has no chords on the first
$n_1$ strings, no chords on the last $n_2$ strings and on the middle
$n$ strings it looks like $D$.

All the operators $\Delta_i,\ve_i,S_{\ell}$ can be extended to
$\cB_n\otimes\Bbb C$.

\section{Non-associative words} A non-associative word on some symbols
is an element of the free non-associative algebra generated by those
symbols. Consider the set of all non-associative
 words on two symbols $+$ and $-$. If $w$ is such a word, different
from $+$ and $-$ and the unit, then $w$
 can be presented in a unique way $w=w_1w_2$, where $w_1,w_2$ are
 non-associative  non-unit words. Define inductively the length
 $l(w)=l(w_1)+l(w_2)$ if $w=w_1w_2$ and $l(+)=l(-)=1$. A
non-associative word can be represented as a sequence of symbols and
parentheses which indicate the order of
 multiplication.

There is a map which transfers each non-associative word into an
associative word by simply forgetting the non-associative structure,
that is, forgetting the parentheses. An associative word is just a
finite sequence of symbols.

If we have a finite sequence of symbols $+,-$, then we can form a
non-associative word by performing the multiplication step by step from
the left. It will be called the standard non-associative word of the
sequence.

Suppose $w_1, w_2$ are  non-associative words. Replacing a symbol in
the word $w_2$ by $w_1$ one gets another word $w$. In such a case we
will call $w_1$ a {\it subword} of $w$, and write $w_1<w$.

\section{q-tangles}

 We  fix an oriented 3-dimensional
 Euclidean space $\R^3$ with coordinates $(x,y,t)$.
{\it A tangle} is a smooth one-dimensional compact  oriented manifold
 $L\subset \R^3$  lying between two horizontal planes $\{t=a\},\{t=b\},
 a<b$ such that all the boundary points are lying on two lines
$\{t=a,y=0\},
\{t=b,y=0\}$, and at every boundary point $L$ is orthogonal to these
two planes. These lines are called the top and the bottom lines
 of the tangle.

{\it A normal vector field} on a tangle $L$ is a smooth vector field on
$L$ which is nowhere tangent to $L$ (and, in particular, is nowhere
zero) and which is given by the vector $(0,-1,0)$ at every boundary
point. {\it A framed tangle} is a tangle enhanced with a normal vector
field. Two framed tangles are isotopic if they can be deformed by  a
1-parameter family of diffeomorphisms into one another within the class
of framed tangles.

  We will consider a
 {\it tangle diagram} as  the projection  onto $\R^2(x,t)$ of tangle in
 generic position. Every the double point is provided with a sign $+$
or $-$ indicating  an over or under crossing.

Two tangle diagrams are equivalent if one can be deformed into another
by using: a) isotopy of $\R^2(x,t)$ preserving every horizontal line
$\{t=\text{const}\}$, b) rescaling of $\R^2(x,t)$, c) translation along
the $t$-axis. We will consider tangle diagrams up to this equivalent
relation.

 Two isotopic framed tangles may project into two non-equivalent tangle
diagrams. But if $T$ is a tangle diagram, then $T$ defines a unique
class of isotopic  {\it framed} tangles $L(T)$: let $L(T)$ be  a tangle
which projects into $T$ and is coincident with $T$ except for a small
neighborhood of the double points, the normal vector at every point of
$L(T)$ is $(0,-1,0)$.

One can assign a symbol $+$ or $-$ to all the boundary points of a
tangle diagram according to  whether the tangent vector at this point directs
downwards or upwards. Then on the top boundary line of a tangle diagram
we have a
 sequence of symbols consisting of $+$ and $-$. Similarly on the bottom
boundary line there is also a sequence of symbols $+$ and $-$.

A {\it q-tangle diagram} $T$ is a tangle diagram enhanced with two
non-associative words $w_b(T)$ and $w_t(T)$ such that when forgetting
about the non-associative structure from $w_t(T)$ (resp. $w_b(T)$) we
get the sequence of symbols on the top (resp. bottom) boundary line.
{\it  A framed q-tangle} $L$  is a framed tangle  enhanced with two
non-associative words $w_b(L)$ and $w_t(L)$ such that when forgetting
about the non-associative structure from $w_t(L)$ (resp. $w_b(L)$) we
get the sequence of symbols on the top (resp. bottom) boundary line.

If $T_1, T_2$ are q-tangle diagrams such that $w_b(T_1)=w_t(T_2)$
we can define the product $T=T_1\times T_2$ by putting $T_1$ above
$T_2$. The product is a q-tangle diagram with $w_t(T)=w_t(T_1),
w_b(T)=w_b(T_2)$.

Every q-tangle diagram can be decomposed  (non-uniquely) as the
product of the following  {\it basic} q-tangle diagrams:

1a)$X_{w,v}^+$ for a non-associative word $w$ one one symbol $+$
containing
a subword $v=++$, the underlying tangle diagram is  in the
following figure

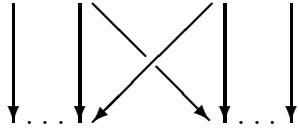
\centerline{ \begin{picture}(105,60) \thicklines
\put(0,45){\vector(0,-1){45}} \put(25,45){\vector(0,-1){45}}
\put(80,45){\vector(0,-1){45}} \put(105,45){\vector(0,-1){45}}
\put(6,0){\makebox(0,0){.}} \put(12,0){\makebox(0,0){.}}
\put(18,0){\makebox(0,0){.}} \put(86,0){\makebox(0,0){.}}
\put(92,0){\makebox(0,0){.}} \put(98,0){\makebox(0,0){.}}
\put(30,45){\line(1,-1){20}} \put(54,21){\vector(1,-1){20}}
\put(75,45){\vector(-1,-1){45}} \end{picture}}

with $w_t=w_b=w$, the two strings of the crossing correspond to two
symbols
 of the word $v$.

1b)$X^-_{w,v}$: the same as $X^+_{w,v}$, only the overcrossing is
replaced by the undercrossing

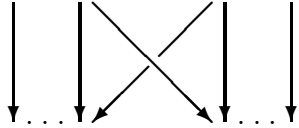
\centerline{ \begin{picture}(105,60) \thicklines
\put(0,45){\vector(0,-1){45}} \put(25,45){\vector(0,-1){45}}
\put(80,45){\vector(0,-1){45}} \put(105,45){\vector(0,-1){45}}
\put(6,0){\makebox(0,0){.}} \put(12,0){\makebox(0,0){.}}
\put(18,0){\makebox(0,0){.}} \put(86,0){\makebox(0,0){.}}
\put(92,0){\makebox(0,0){.}} \put(98,0){\makebox(0,0){.}}
\put(30,45){\vector(1,-1){45}} \put(51,21){\vector(-1,-1){21}}
\put(75,45){\line(-1,-1){20}} \end{picture}}

2a)$\cup_{w,v}$ with $v=(+-)<w$, all the symbols in $w$ outside $v$
are $+$.
 The underlying tangle is

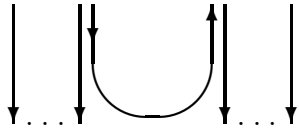
\centerline{ \begin{picture}(105,60) \thicklines
\put(0,45){\vector(0,-1){45}} \put(25,45){\vector(0,-1){45}}
\put(80,45){\vector(0,-1){45}} \put(105,45){\vector(0,-1){45}}
\put(6,0){\makebox(0,0){.}} \put(12,0){\makebox(0,0){.}}
\put(18,0){\makebox(0,0){.}} \put(86,0){\makebox(0,0){.}}
\put(92,0){\makebox(0,0){.}} \put(98,0){\makebox(0,0){.}}
\put(30,45){\vector(0,-1){15}} \put(52.5,30){\oval(45,55)[b]}
\put(75,30){\vector(0,1){15}} \end{picture}} Here $w_t=w$,
$w_b$ is obtained from $w$ by removing $v$.

2b)$\cap_{w,v}$ with $v=(-+)<w$, all the symbols in $w$ outside
$v$ are $+$.
The underlying tangle is

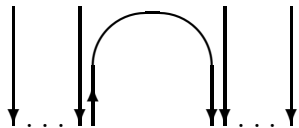
\centerline{ \begin{picture}(105,60) \thicklines
\put(0,45){\vector(0,-1){45}} \put(25,45){\vector(0,-1){45}}
\put(80,45){\vector(0,-1){45}} \put(105,45){\vector(0,-1){45}}
\put(6,0){\makebox(0,0){.}} \put(12,0){\makebox(0,0){.}}
\put(18,0){\makebox(0,0){.}} \put(86,0){\makebox(0,0){.}}
\put(92,0){\makebox(0,0){.}} \put(98,0){\makebox(0,0){.}}
\put(75,15){\vector(0,-1){15}} \put(52.5,15){\oval(45,55)[t]}
\put(30,0){\vector(0,1){15}} \end{picture}}

Here $w_b=w$, $w_t$ is obtained from $w$ by removing $v$.

3a)$T^+_{w_1w_2w_3,w}$ where $w_1,w_2,w_3,w$ are non-associative words
on one symbol $+$, and $((w_1w_2)w_3)$ is a subword of $w$. The
underlying tangle
 diagram is trivial, consisting of $l(w)$ parallel lines, all are
directed downwards, and $w_b(T^+_{w_1w_2w_3,w})=w$ while
$w_t(T^+_{w_1w_2w_3,w})$ is obtained from $w$ by substituting
$((w_1w_2)w_3)$ by $(w_1(w_2w_3))$.

3b)$T^-_{w_1w_2w_3,w}$ where $w_1,w_2,w_3,w$ are non-associative words
on one symbol $+$, and $((w_1w_2)w_3)$  is a subword of $w$. The
underlying tangle diagram is trivial, consisting of $l(w)$ parallel
lines, all are directed downwards, and $w_t(T^-_{w_1w_2w_3,w})=w$
while $w_b(T^-_{w_1w_2w_3,w})$ is obtained from $w$ by substituting
$((w_1w_2)w_3)$ by $(w_1(w_2w_3))$.

4)All the above listed q-tangle diagrams with reversed orientations
on some strings and  the corresponding  change of signs of the boundary
points.

\section{ The Drinfeld associator} Let $M$ be the algebra over
$\Bbb C$ of
 all formal  series on two non-commutative, associative symbols
 $A,B$.  Consider a function $G:(0,1)\rightarrow M$ satisfying the
following Knizhnik-Zamolodchikov equation $$\frac{\text d}{\text
dt}G=\frac{1}{2\pi\sqrt{-1}}(\frac{A}{t}+\frac{B}{t-1})G.$$ Let
$G_\lambda(A,B)$ be the value at $t=1-\lambda$ of
 the solution to this equation which takes  the value 1 at $t=\lambda$.
 It
can be proved that the following  limit exists
$$\varphi(A,B)=\lim_{\lambda\rightarrow 0}\lambda^{-B}G_\lambda
\lambda^A.$$
It can be written in the form $$1+\sum_{W} f_WW,$$ where
the summation is over all the associative words on two symbols $A$ and
$B$.
The coefficients $f_W$ are highly transcendent and are computed in
\cite{Drin2,Le-Mu2}. Each $f_W$ is the  sum of a finite number
of numbers of type
$$\frac{1}{(2\pi \sqrt{-1})^{i_1+\dots+i_k}}\zeta(i_1,\dots,i_k),$$
 where $$\zeta(i_1,\ldots,i_k)=\sum_{n_1<\ldots<n_k\in\N}\frac{1}
{n_1^{i_1}\cdots
 n_k^{i_k}}$$ with natural numbers $i_1,\ldots,i_k$,$i_k\ge2$. These
numbers have recently gained
 much attention among number theorists (see \cite{Zagier}).

Denote by $\Omega_{ij}$  the chord diagram in $\cB_n$
 with one chord connecting the $i$-th and the $j$-th strings.  Let
$\Phi_{KZ}=\varphi(\Omega_{12},\Omega_{23})\in\cB_3\otimes\Bbb C$.
This
element is called  the KZ Drinfeld
 associator.  It is a solution of the following equations (for
 a proof see \cite{Drin1,Drin2}):  \begin{equation} (\id\otimes \id
\otimes\Delta)(\Phi)\times(\Delta\otimes\id\otimes \id )(\Phi)=
(1\otimes \Phi)\times(\id\otimes \Delta\otimes\id
)(\Phi)\times(\Phi\otimes 1),\tag{A1} \end{equation} \begin{equation}
(\Delta\otimes\id)(R)=\Phi^{312}\times
R^{13}\times(\Phi^{132})^{-1}\times R^{23}\times\Phi, \tag{A2a}
\end{equation} \begin{equation}
(\id\otimes\Delta)(R)=(\Phi^{231})^{-1}\times R^{13}\times
\Phi^{213}\times R^{12}\times\Phi^{-1}, \tag{A2b} \end{equation}
\begin{equation}\Phi^{-1}=\Phi^{321},\tag{A3}\end{equation}
\begin{equation}\ve_1(\Phi)=\ve_2(\Phi)=\ve_3(\Phi)=1.\tag{A4}
\end{equation}
 Here $\Phi^{ijk}$ is the element of
$\cB_3\otimes \Bbb C$ obtained from $\Phi$ by  permuting the strings:
the first to the $i$-th, the second to the $j$-th, the third to the
$k$-th and  $ R^{ij}=\exp(\Omega_{ij}/2)$. Equation (A1) holds in
$\cB_4\otimes\C$, equations (A2a,A2b,A3) hold in $\cB_3\otimes\C$,
equation (A4)  in $\cB_2\otimes \C$.

\noindent{\bf Remark:} (A2b) follows from the other identities
in (A1-A4).

Besides, $\Delta$, $\Phi_{KZ} $ and
$R=\exp(\Omega_{12}/2)\in\cB_2\otimes\Bbb C$  satisfy:
\begin{equation}
(\id\otimes\Delta)\Delta(a)=\Phi((\Delta\otimes\id)\Delta(a))
\Phi^{-1},\tag{B1}
\end{equation}
\begin{equation} (\varepsilon\otimes\id)_{\text
o}\Delta=\id=(\id\otimes\varepsilon)_{\text
o}\Delta,\tag{B2}\end{equation}
\begin{equation}\Delta(a)=R\Delta(a)R^{-1}. \tag{B3}\end{equation}

The first follows from (1) and theorem 1 for any $\Phi$, the second is
trivial,  the third follows from theorem 1.

 Every solution $\Phi$ of (A1-A4) is called an associator.
 Theorem $A''$
 of \cite{Drin2} asserts that there is  an associator
$\Phi_{\Bbb Q}$  with rational  coefficients, i.e $\Phi_{\Bbb
Q}\in\cB_3$.
\section{ The representation of  framed q-tangles}

 Every tangle diagram $T$ defines a framed tangle $L(T)$, and every
 framed tangle  $K$ is $L(T)$ for some tangle diagram.

Suppose $T$ is a q-tangle diagram. Then $L(T)$ is a framed q-tangle.
 Regarding $L(T)$ as a 1-dimensional manifold, we can define the vector
 space $\cA(L(T))$, which we will abbreviate as $\cA(T)$. This vector
 space depends
  only on the underlying tangle  diagram of $T$ but not on the words
 $w_b$ and $w_t$.

 If $D_i\in \cA(T_i),i=1,2$ and $T=T_1\times T_2$ then we can define
 $D_1\times D_2\in\cA(T)$ in the obvious way, just putting $D_1$ above
 $D_2$.

 We will define a mapping $T\rightarrow \Zf(T)\in\cA(T)\otimes \Bbb C$
 for any q-tangle diagram such that if $T=T_1\times T_2$ then
 $\Zf(T)=\Zf(T_1)\times \Zf(T_2)$.  Such a
 map is uniquely defined by the values of special q-tangle diagrams
 listed in the previous section.

 Define

\noindent (D1a) \quad $\Zf(X^+_{w,v})=\exp(\Omega/2):=
1+\Omega/2+\frac{1}{2!}(\Omega/2)^2+\cdots,$ \\
 where $\Omega^n$ is the chord diagram in $\cA(X_{w,v}^+)$ with $n$
 chords,
 each is parallel to the horizontal line and connects the two strings
 that form the double point of $X_{w,v}^+$.

\noindent (D1b)\quad $\Zf(X^-_{w,v}):=\exp(-\Omega/2)$.

\noindent (D2a)\quad $\Zf(\cup_{w,v})$ is the chord diagram in
$\cA(\cup_{w,v})$ without any
 chord.

\noindent (D2b)\quad $\Zf(\cap_{w,v})$ is the chord diagram in
$\cA(\cap_{w,v})$ without any
 chord.

\noindent (D3)\quad For a q-tangle diagram $T^{\pm}_{w_1w_2w_3,w}$
let $\#l$ and $\#r$ be respectively the number of strings  (in the
underlying tangle diagram) left
 and right  of the block of strings which form the word
 $((w_1w_2)w_3)$. Define

 $\Zf(T^+_{w_1w_2w_3,w})=1^{\otimes\#l}\otimes[(\Delta^{l(w_1)-1}
\otimes\Delta^{l(w_2)-1}
 \otimes\Delta^{l(w_3)-1})\Phi_{KZ}]\otimes 1^{\otimes\#r}$

 $\Zf(T^-_{w_1w_2w_3,w})=1^{\otimes\#l}\otimes[(\Delta^{l(w_1)-1}
\otimes\Delta^{l(w_2)-1}
 \otimes\Delta^{l(w_3)-1})(\Phi_{KZ})^{-1}]\otimes 1^{\otimes\#r}$

\noindent (D4)\quad If $T'$ is a q-tangle diagram obtained from $T$
by reversing the orientation
 of some components $\ell_1,\ldots,\ell_k$, then $\Zf(T')$ is obtained
 from $\Zf(T)$ by successively applying the mappings
 $S_{\ell_1},\ldots,S_{\ell_k} $. The result does not depend on the
 order of these mappings.

 \noindent{\bf Theorem 2:}     {\it

 1. The mapping $T\rightarrow \Zf(T)$ is  well-defined: it does not
 depend on the  decomposition of  a q-tangle diagram into basic
 q-tangle diagrams.

2. Let $\phi=\Zf(U)\in \cA\otimes\Bbb C$, where $U$ is tangle diagram
in the following figure

 \centerline{\begin{picture}(30,40)(-15,0)\thicklines
       \put(0,20){\oval(30,40)[b]} \put(0,20){\oval(10,20)[b]}
       \put(10,20){\oval(10,20)[t]}\put(-10,20){\oval(10,20)[t]}
\end{picture}}

Then $$Z_f(\begin{picture}(10,20)(0,7.5)
\thicklines
\put(0,0){\line(1,1){10}}
\put(0,10){\line(0,1){10}}
\put(2.5,10){\oval(5,5)[b]}
\put(7.5,10){\oval(5,5)[t]}
\end{picture})=\phi.\Zf(\begin{picture}(10,20)(0,7.5)
\thicklines
\put(5,0){\line(0,1){20}}
\end{picture}
).$$
The right hand side is the connected sum of $\phi$ and $\Zf$ along the
 indicated component.

3. Suppose the coordinate function $t$ on the $i$-th  component of $T$
  has $s_i$ maximal points.  Define
$$\hat{\Zf}(T)=(\phi^{-s_1}\otimes\cdots\otimes\phi^{-s_k}).\Zf(T)$$
where  the right hand side is the element obtained by  successively
 taking the connected sum of  $\phi^{-s_i}$ and  $\Zf(T)$ along the
$i$-th component.  If two q-tangle diagrams $T_1,T_2$ define isotopic
 framed q-tangles, $L(T_1)=L(T_2)$, then $\hat Z_f(T_1)=\hat Z_f(T_2)$,
 hence  $\hat{\Zf}$ is an  isotopy invariant of framed
q-tangles.      }

In particular, $\hat Z_f$ is an  isotopy invariant of {\it framed
oriented links} regarded as framed q-tangles without boundary points.

There are at least two ways to prove Theorem 2. In the first  which is
more algebraic, we use MacLane's coherence theorem to reduce
 the category of q-tangles to the
category of tangles and then verify that $\hat Z_f$ does not change
under certain local moves (see the definition of the local moves
in \cite{Re-Tu,Coste}). Similar proofs are
sketched in \cite{Car,Piu}. In the second which is more analytical (see
\cite{Le-Mu3}),  we first define the regularized Kontsevich integral
for framed  oriented tangles, then we prove that the value $\Zf$ of a
q-tangle is the limit (in some sense) of the regularized Kontsevich
integrals. In this approach we can avoid MacLane's cohenrence theorem
and veryfying the invariance under local moves. The second proof also
show the relation between $\hat Z_f$ and the original Kontsevich
integral (see Theorem 6 below).

\noindent{\bf Remark:} We have chosen the normalization in which
$\hat Z_f$ of the unframed trivial knot is $\phi^{-1}$, of the
 empty knot is $1$.

\Thm{3:} {\it Let $\omega$ be the  unique element of $\cA$ with one
chord. Then a change in  framing results on $\hat{\Zf}$ by multiplying
by $\exp(\omega/2)$:
$$e^{-\omega/2}.\hat{\Zf}(\ {\begin{picture}(30,40)(0,10)
\put(22,0){\line(-1,1){10}} \put(0,0){\line(1,1){22}}
\put(0,22){\line(1,-1){10}} \put(30,0){\line(0,1){22}}
\put(0,0){\line(0,-1){15}} \put(0,22){\line(0,1){15}}
\put(26,22){\oval(8,10)[t]} \put(26,0){\oval(8,10)[b]} \end{picture}}\
)=\hat{\Zf}(\ {\begin{picture}(10,40)(0,22) \put(5,0){\line(0,1){52}}
\end{picture}}\ )=e^{\omega/2}.\hat{\Zf}(\ {\begin{picture}(30,40)(0,10)
\put(22,0){\line(-1,1){22}} \put(30,0){\line(0,1){22}}
 \put(0,0){\line(1,1){10}} \put(22,22){\line(-1,-1){10}}
\put(0,0){\line(0,-1){15}} \put(0,22){\line(0,1){15}}
\put(26,22){\oval(8,10)[t]} \put(26,0){\oval(8,10)[b]} \end{picture}}\
)$$ }

This can be proved easily by moving the minimum point to the left then
 using the representations of $q$-tangles. The invariant $\hat{\Zf}$
is coincident with the invariant introduced
in \cite{Le-Mu2,Le-Mu3} of framed links. There it was constructed by
modifying the original Kontsevich integral.

$\hat{Z}_f$ is called a  universal Vassiliev-Kontsevich invariant
 of framed oriented  links. As in
\cite{Bar}, it is easy to prove that $\hat{Z}_f(K_1)=\hat{Z}_f(K_2)$ if
and only if all the invariants of finite type are the same for framed
links $K_1$ and $K_2$. This means $\hat{Z}_f$ is as powerful as the set
of all invariants of finite type, in particular it dominates all
invariants coming from $R$-matrices which are  deformations of
 identity, as in \cite{Tu,Re-Tu}.

Let $\ell$ be a component of  a one-dimensional compact manifold $X$
and $X'$
 be obtained from $X$ by replacing  $\ell$ by two copies of $\ell$.  In
 a similar manner as in \S1 one can define the operator
 $\Delta_\ell:\cA(X)\rightarrow\cA(X')$.

\Thm{4:} {\it Suppose $L$ is a  framed oriented link with a distinguished
component $\ell$ ,
 $L'$ is obtained from $L$ by replacing $\ell$ by two its
 parallels, close to $\ell$, $L''$ is obtained from $L$ by reversing
 the orientation of $\ell$. Then
$${\hat Z_f}(L')=\Delta_\ell({\hat Z_f}(L)).$$
$${\hat Z_f}(L'')=S_\ell({\hat Z_f}(L)).$$}
The second identity follows trivially from the definition of $\hat Z_f$.
 The first is more difficult and can be proved by analyzing the
parallel of the basic $q$-tangles. Note that the chosen normalization
of $\hat Z_f$ plays important role in the second identity. The
 formula for  the parallel version of $\Zf$ (not $\hat Z_f$) would
 require an additional factor.
 Applying this identity to the unknot we get a beautiful formula
 $\Delta(\phi)=\phi\otimes\phi$.

\Thm{5:} {\it Suppose $L_1, L_2$ are framed  links with
distinguished
 components and $L$ is the connected link along the distinguished
 components. Then $$\hat{\Zf}(L)=\phi.(\hat{\Zf}(L_1)).(\hat{\Zf}(L_2))$$
}
The right hand side is the connected sum of $\phi$, $\hat{\Zf}(L_1)$ and
$\hat{\Zf}(L_2)$ along the distinguished components.

 Theorem  5 can be proved easily using the representation of
$q$-tangles, or using the integral formula in \cite{Le-Mu3}.

\section{The Kontsevich integral} Let $\cA_0$ be the  vector space  over
$\Q$ (rational numbers) spanned by all chord diagram with support being
a circle, subject to the 4-term relation and the following framing
independence relation:

$$\raisebox{-8pt}{\begin{picture}(40,20)(-20,-20)
\put(-10,-6){\dashbox{2}(20,0){}} \thicklines
\put(0,0){\oval(20,40)[b]} \end{picture}}=0$$

 In other words, $\cA_0=\cA/(\omega\cA)$. With respect to  connected
 sum, $\cA_0$ is a commutative algebra. There is a natural projection
$p:\cA\rightarrow \cA_0$.

Let $K$ be the image of an embedding of the  oriented circle
 into  $\R^3$ lying between two horizontal planes $\{t=t_{min}\},
\{t=t_{max}\}$. We will consider the 2-dimensional plane $(x,y)$
as the complex plane with coordinate $z=x+y\sqrt{-1}$. The Kontsevich
integral of $K$ is defined  as an element of $\cA_0$ by

 $$Z(T)=\sum_{m=0}^\infty \frac{1}{(2\pi\sqrt{-1})^m}\int_{t_{min}
<t_1<\dots<t_m<t_{max}}\sum_
{P}(-1)^{\#P\!\downarrow}\wedge\frac{dz_i-
dz^{\prime}_
i}{z_i-z^{\prime}_i}D_P\in\cA_0$$

 where for fixed $t_{min}
<t_1<t_2
<\dots<t_m<
t_{max}$ the object $P$ is a choice of  unordered pairs of  distinct
 points $z_i,z_i'$ lying in $K\cap\{t=t_i\}$ for $i=1\dots,m$,
 the summation
is over all such choices,
 $D_P$ is the chord diagram  in $\cA_0$ obtained by
connecting pairs $z_i,z^{\prime}_i$ by dashed lines,
 $\#P\negthickspace\downarrow$ is the number of $z_i,z_i'$ at
 which the orientation of $K$
is downwards. Here we regard $z_i,z_i'$ as functions of $t_i$
(for more details on the Kontsevich integral see \cite{Bar}).

The integral $Z(K)$ is not an isotopy invariant. Let $\gamma=p(\phi)$.
 Kontsevich proved that
$\hat Z(K):=\gamma^{-s}.Z(K)$, where $s$ is the number of maximum
 points of the coordinate function $t$ on $K$, is an isotopy
 invariant of (unframed) oriented knots. Note that in \cite{Bar}
instead of $\hat Z$ another normalization $\tilde Z=\gamma.\hat Z$
is used.
 This invariants is as powerful
 as the set of all invariants of finite type. The relation between
$\hat Z_f$ and the Kontsevich integral is explained in the following

\Thm{6:} {\it For a framed oriented knot $K$
 $$p( \hat{\Zf}(K))=\hat Z(K).$$}
This  theorem  and the trivial generalization for links are proved
 in \cite{Le-Mu3}. Knowing $\tilde Z(K)\in\cA_0$ one can
 also compute $\hat Z_f(K)$ (see \cite{Le-Mu3}).

\section{Symmetric Twisting} An element $D\in\cB_2\otimes \C$ is called
{\it symmetric} if $D^{21}=D$, where $D^{21}$ is obtained from $D$ by
permuting the two strings of the support. Let $F=1+F_1+F_2+\cdots
\in\cB_2\otimes\Bbb C$, where $F_n$ is the homogeneous part of grading
$n$. Suppose

\noindent (T1) for $n\ge 1$ every chord diagram in $F_n$ has
vertices on both
strings, i.e. $F_n\in(ker\ve_1\cap ker\ve_2)$.

\noindent (T2) $F$ is symmetric.

Then there exist $F^{-1}$ in $\cB_2\otimes \C$ satisfying (T1,T2).
\newcommand{\cF}{{\cal F}}

If $\Phi$ is an element of $\cB_3\otimes\C$ then
 \begin{equation}\tilde\Phi=[1\otimes
 F][(id\otimes\Delta)F]\Phi[\Delta\otimes id)(F^{-1}][F^{-1}\otimes
 1]\end{equation} is said to be obtained from $\Phi$ by a twist using
$F$.

Note that the first two terms  in the right hand side of (2) are
commutative with each other, as are the last two terms. If
$\Phi\in\cB_3\otimes\Bbb C$ is a solution of (A1-A4) then it is not
difficult to check that $\tilde\Phi$ is also a solution of (A1-A4).

For a non-associative word $w$ on one symbol $+$ define
$\cF_w\in\cB_{l(w)}$ by
 induction as follows. Let $\cF_{\emptyset}=1\in\Q$, $\cF_+=1\in\cB_1$,
 $\cF_{++}=F\in\cB_2\otimes\C$. For $w=w_1w_2$ let
$$\cF_w=[\cF_{w_1}\otimes 1^{\otimes l(w_2)}][1^{\otimes l(w_1)}\otimes
 \cF_{w_2}][(\Delta^{l(w_1)-1}\otimes\Delta^{l(w_2)-1})F]$$

Then (2) implies that $\tilde\Phi=\cF_{+(++)}\Phi(\cF_{(++)+})^{-1}$.

\newcommand{\ZFf}{Z^F_f}
 Fix $F\in\cB_2\otimes\C$ satisfying (T1,T2). Consider a new mapping
 $T\rightarrow \ZFf(T)$ defined for q-tangle diagrams by the same rules
 (D1-D4) for  $\Zf$, only  replacing the values listed in \S 3  for
 basic
q-tangle diagrams by:
$$\ZFf(X_{w,v}^+)=\Zf(X_{w,v}^+),$$ $$\ZFf(X_{w,v}^-)=
\Zf(X_{w,v}^-),$$
\begin{equation}\ZFf(\cup_{w,v})=[1^{\otimes m}\otimes F\otimes
1^{\otimes n}]\times\Zf (\cup_{w,v})\tag{$D2a'$}\end{equation}
\begin{equation}\ZFf(\cap_{w,v})=\Zf(\cap_{w,v})\times[1^{\otimes
m}\otimes F^{-1}\otimes 1^{\otimes n}]\tag{$D2b'$}\end{equation}
The values of $\ZFf(T^\pm_{w_1w_2w_3,w})$ are defined by the same
formulas  as in
(D3), only $\Phi_{KZ}$ is replaced by $\tilde\Phi$ obtained from
$\Phi_{KZ}$ by  a twist using $F$.

\Thm{7:} {\it The map $\ZFf$ is  well-defined and for every
q-tangle  diagram $T$
\begin{equation}\label{twist}\ZFf(T)=\cF_{w_t(T)}\Zf(T)
[\cF_{w_b(T)}]^{-1}
.\end{equation} In particular, $\ZFf(L)=\Zf(L)$ for
 any tangle diagram
$L$ without boundary points.}

\Pf We need only to check identity (\ref{twist}). It's sufficient  to
check  for the cases when $T$ are basic q-tangle diagrams. These
cases follows trivially from the definition.\qed

Note that if $\tilde\Phi$ has rational coefficients, i.e. if
$\tilde\Phi\in\cB_3$, then from the definition, the invariant $\ZFf$ of
a framed {\it link} (not framed q-tangle) has rational coefficients,
$\ZFf(K)\in\cA(K)$. Although  the coefficients of $F$ may be irrational
and in $(D2a',D2b')$ the elements $F,F^{-1}$
are involved, they appear in pairs which annihilate each other in every
link diagram.

\noindent{\bf Remark:} In \cite{Drin1,Drin2} Drinfeld defined twists
for  quasi-triangular quasi-Hopf algebras. Here we adapt a similar
definition for the series of algebras $\cB_n$ which play the role of a
{\it single} quasi-triangular quasi-Hopf algebra. If we use the
representation    of section 10 below then we get a quasi-triangular
quasi-Hopf algebra, and the construction of twists here corresponds
only to Drinfeld's twist which does not change the co-multiplication.
If $F$ is not symmetric, then $\Delta$ is replaced by
$\tilde\Delta=F^{21}\Delta F^{-1}=(F^{21}F^{-1})\Delta$.

\section{Uniqueness and rationality of the universal
 Vassiliev-Kontsevich
invariant}

\Thm{8:} {\it  If $\Phi,\Phi'\in(\cB_3\otimes \Bbb C) $ satisfy
(A1-A4), then $\Phi$ is obtained from $\Phi'$ by  a twist using
$F\in\cB_2\otimes\Bbb C$ satisfying (T1,T2).}

As a corollary, from every solution $\Phi$ of (A1-A4) we can construct
an invariant of framed q-tangles. All such invariants, when restrict
to the sets of framed  oriented links, are the same and  contain  all
invariants of framed  oriented links of finite type. By theorem $A''$ of
\cite{Drin2} there is a solution $\Phi_{\Bbb Q}$  with rational
coefficients,  thus we get

\noindent{\bf Corollary}: {\it The universal Vassiliev-Kontsevich
 invariant of framed links
has rational coefficients in the sense that $\hat Z_f(L)$ belongs to
$\cA(L)$ for every framed link $L$. The Kontsevich integral of a knot
has rational coefficient in the sense that $\tilde{Z}(K)\in\cA_0$.}

\noindent {\bf Proof} of Theorem 8:  Let
$$\Phi=1+\Phi_1+\dots+\Phi_n+\cdots$$
$$\Phi'=1+\Phi'_1+\dots+\Phi'_n+\cdots$$ Here $\Phi_n$, $\Phi'_n$ are
the  homogeneous part of grading $n$. Suppose
 we already have $\Phi_i=\Phi'_i$ for $0\le i\le k-1$. Put
 $\psi=\Phi_k-\Phi'_k$.

Comparing the $k$-grading parts of (A1-A4) for $\Phi,\Phi'$ we get:
\begin{equation} \text d\psi=0,\tag{C1}\end{equation} \begin{equation}
\psi-\psi^{132}-\psi^{213}=0,\tag{C2}\end{equation} \begin{equation}
\psi^{321}=-\psi,\tag{C3}\end{equation} \begin{equation} \ve_1(\psi)=
\ve_2(\psi)=\ve_3(\psi)=0,\tag{C4}\end{equation} where $\text
d:\cB_n\to\cB_{n+1}$ is the mapping:  $$ \text d(a)=1\otimes
a-\Delta_1(a)+\Delta_2(a)-\dots+(-1)^n\Delta_n(a)+(-1)^{n+1}a\otimes
1.$$ We extend $d$ to $\cB_n\otimes \Bbb C$.

\Pro{3:} {\it If $\psi\in\cB_3\otimes \Bbb C$ of grading $k$ and
satisfying (C1-C4) then there is a symmetric element $f\in\cB_2\otimes
\Bbb C$ of grading $k$  such that $\text d(f)=\psi$;
$\varepsilon_1(f)=\varepsilon_2(f)=0$.}

Suppose for the moment that this is true. Pick $f$ as in this
proposition. Then  one can check immediately that the twist by $F=1+f$
transfers $\Phi$ to $\tilde{\Phi}$ with  $\tilde{\Phi}_i=\Phi'_i$ for
$0\le i\le k$.

Continue the process we can find a  element $F\in\cB_2\otimes \Bbb C$
satisfying (T1,T2) which transfers $\Phi$ into $\Phi'$.

There remains Proposition 3 to prove.

\section{Proof of Proposition 3} \subsection{Other realizations of
{$\cB_n$}}

{\it A Chinese character} (\cite{Bar}) is a graph whose vertices
are either
trivalent and oriented or univalent. Here an orientation of a
 trivalent
vertex is just a cyclic order of the three edges incident to this
vertex. The trivalent vertices are called {\it internal}, the univalent
vertices are called {\it external}. The edges of the graph will be
represented by dashed lines on the plane. By convention all the
orientations in figures  are counterclockwise for  Chinese characters.

An {\it $n$-marked Chinese character} $C$ is a Chinese character  with
at least one external vertex in each connected component,
where in addition the external
vertices are partitioned into $n$  labeled sets $\Theta_1(C),
\dots,\Theta_n(C)$.

Let $\cC_n$ be the vector space over $\Bbb Q$ spanned by all
$n$-marked
Chinese characters subject to the antisymmetry vertex and IHX
identities (see also \cite{Bar}):

(1) the antisymmetry of internal vertices:
\vskip 2cm

(2) The IHX identity
\vskip 2cm

Let us define  linear mappings $\Delta_i:\cC_n\to\cC_{n+1}$ and
$\ve_i:\cC_n\to\cC_{n-1}$. Suppose the set $\Theta_i(C)$ of an
$n$-marked  Chinese
character $C\in\cC_n$ contains exactly $m$ vertices. There are $2^m$
ways of partition $\Theta_i(C)$ into an ordered pair of subsets.
 For each
such partition $q$ let $D_q$ be the $(n+1)$-marked Chinese character
with the same underlying Chinese character as $C$,
$\Theta_j(D_q)=\Theta_j(C)$ if
$j<i$, $\Theta_j(D_q)=\Theta_{j-1}(C)$ if $j\ge i+2$,
while $\Theta_i(D_q),
\Theta_{i+1}(D_q)$ are two subsets of $\Theta_i(C)$
 corresponding to the
partition $q$.  Define $\Delta_i(C)$ as the sum of all  $2^m$\quad
$(n+1)$-marked Chinese characters $D_q$.

If $\Theta_i(C)\not=\emptyset$ define $\ve_i(C)=0$. Otherwise define
$\ve(C)$ as the $(n-1)$-marked Chinese   character with the same
underlying Chinese character and $\Theta_j(\ve_i(C))=\Theta_j(C)$
 if $j<i$,
$\Theta_j(\ve_i(C))=\Theta_{j+1}(C)$ if $j\ge i$.

 {\it The  {$\Bbb Z^n$-grading}} of an $n$-marked Chinese character $C$
 is the tuple $(k_1,\dots,k_n)$ of integers, where $k_i$ is the number
 of elements of $\Theta_i(C)$. The number $\sum_{i=1}^nk_i$ is called the
 $\Bbb Z$-grading of $C$.  Note that all the mappings $\Delta_i,\ve_i$
respect   the  $\Z$-grading.

We define the linear mapping $d:\cC_n\to\cC_{n+1}$ by $$d(C)=1\otimes
C-\Delta_1(C)+\Delta_2(C)-\dots+(-1)^n\Delta_n(C)+(-1)^{n+1}C\otimes
1.$$ Here $1\otimes C$ and $C\otimes 1$ are the $(n+1)$-marked Chinese
characters gotten from modifying the marking on $C$ by setting
$\Theta_1(1\otimes
C)=\emptyset$, $\Theta_j(1\otimes C)=\Theta_{j-1}(C)$ for $2\le
 j\le n+1$,
$\Theta_{n+1}(1\otimes C)=\emptyset$, $\Theta_j(1\otimes C)
=\Theta_{j}(C)$ for $1\le
j\le n$.

Now we define a linear mapping $\chi:\cC_n\to\cB_n$ as follows.
 First we define $\chi'(C)$ for an $n$-marked Chinese character
$C$ of $\Z^n$-grading $(k_1,\dots,k_n)$. There are
$k_i!$ ways to put vertices from $\Theta_i(C)$ on the $i$-th
string and each of  the $k_1!\dots k_n!$ possibilities gives us an
element of $\cB_n$. Sum up all such elements we get $\chi'(C)$.

Now we use the following STU relation
\vskip 3cm

to convert every diagram appearing  in $\chi'(C)$ into chord diagram,
by that way from $\chi'(C)$ we get $\chi(C)$.

\Thm{9:} {\it  The linear  mapping  $\chi$ is well-defined and is an
isomorphism between vector spaces $\cC_n$ and $\cB_n$ commuting  with
all the operators $\Delta_i,\ve_i$.}

\noindent Remark: $\chi$, however, does not preserve gradings.

The proof for the case $n=1$ is presented  in
\cite[Theorems 6 \& 8]{Bar}. This proof does not concern the support
of chord diagrams except for the first step of the induction which
is trivial in case $n\ge 1$ (see also \cite{Bar2}).

Consider the following subspaces $\cG_n$ of $\cC_n\otimes\C$,
$\cG_n=\cap_{i=1}^n \text{ker} (\ve_i)$. It can be checked that
$d(\cG_n)\subset \cG_{n+1}$. We will now study the homology of the
following chain complex:  \begin{equation}
\label{complex}0\overset{d}{\to}\cG_1\overset{d}
\to\dots\overset{d}\to\cG_n\overset{d}\to\cG_{n+1}
\overset{d}\to\cdots.\tag{*}\end{equation}
Note that $d$ preserves the $\Z$-grading, hence it suffices to
consider the part of $\Z$-grading $m$ of the complex.
 \begin{equation}
\label{complex-m}0\overset{d}\to\cG_1^m\overset{d}\to\dots
\overset{d}\to\cG_n^m\overset{d}\to\cG_{n+1}^m\overset{d}
\to\cdots,\tag{$*_m$}\end{equation}

where $\cG_n^m$ is the homogeneous part of $\Z$-grading $m$  of
$\cG_n$. We will find a geometric interpretation of this complex.

\subsection{A simplicial complex of the cube} Let $I^m$ be the
$m$-dimensional cube, $$I^m=\{\sum_{i=1}^m\lambda_iv_i \quad |\quad
\lambda_i\in[0,1]\}$$ where $v_1,\dots,v_m$ form a base of $\R^m$. We
partition $I_m$ into $m!$\quad  $m$-simplexes: a permutation
$(i_1,\dots,i_m)$ of $(1,\dots,m)$ gives rise to the $m$-simplex which
is the convex hull of $m+1$ points $0,v_{i_1},v_{i_1}+v_{i_2},\dots,
v_{i_1}+\dots+v_{i_m}$. This turns $I^m$ into a simplicial complex,
denoted by $C(I^m)$. The space $C_k(I^m)$ is the vector space over $\C$
spanned by all the $k$-facets of all $m!$ above $m$-simplexes. The
boundary $\partial(I^m)$ is a  simplicial subcomplex. The space
$C_k(\partial(I^m))$ is spanned by all $k$-facets which lie entirely
in $\partial(I^m)$.

Let $C_k$ be the vector space over $\C$ spanned by all tuples
$(\theta_1,\dots,\theta_k)$ which are partitions of the set
 $\{1,2,\dots,m\}$,
each $\theta_i$ non-empty. Define $\partial:C_k\to C_{k-1}$ by
$$\partial(\theta_1,\theta_2,\dots,\theta_k)=(\theta_1\cup
\theta_2,\theta_3,\dots,\theta_k)-(\theta_1,\theta_2\cup
\theta_3,\dots,\theta_k)+\dots+(-1)^{k-1}(\theta_1,\dots,
\theta_{k-1}\cup \theta_k).$$ Then the
chain complex $(C_*,\partial)$ is isomorphic to the quotient complex
$C(I^m)/C(\partial(I^m))$. In fact, the mapping which sends
 $(\theta_1,\dots,\theta_k)$  to the
$k$-simplex with vertices
$0,v_{\theta_1},v_{\theta_1}+v_{\theta_2},\dots,v_{\theta_1}+
\dots+v_{\theta_k}$ is an
isomorphism between these two complexes, where $v_\theta=
\sum_{j\in \theta}v_j$.

Let $E_m$ be the dual  chain complex of $(C_*,\partial)$,
 $E_m=(C^*,\delta)$.
Using the above base of $C_k$, we can identify $C_k^*$ with
$C_k$ with
the same base. Then the co-boundary  $\delta$ can be written
explicitly as
$$\delta(\theta_1,\theta_2,\dots,\theta_k)=(\delta(\theta_1),
\theta_2,\dots,\theta_k)-(\theta_1,\delta(\theta_2),\dots,
\theta_k)+\dots+(-1)^{k-1}(\theta_1,\theta_2,\dots,\delta(\theta_k)),$$
where for a non-empty subset $\theta$ of $ \{1,2,\dots,m\}$ we set
$\delta(\theta)=\sum(\theta',\theta'')$, the summation  is over
all possible partitions of
$\theta$ into an  ordered pair of non-empty subsets.

\Pro {4:} {\it The homology of the chain  complex $E_m$ is
 given by $H_m(E_m)=\C$,
$H_i(E_m)=0$ for $0\le i\le m-1$.}

This follows from the fact that the homology of $E_m$ is the
 reduced cohomology
of $I^m/\partial(I^m)$.

Since every tuple $(\theta_1,\dots,\theta_k)\in C_k$ is a partition of
$\{1,2,\dots,m\}$, the symmetric group $S_m$ acts naturally on $C_k$.
In the simplicial complex $C(I^m)$ this corresponds to the action:
$(v_1,\dots,v_m)\to (v_{\sigma(1)},\dots,v_{\sigma(m)})$ for
$\sigma\in S_m$. On (co)homology the action is trivial.

\Pro {5:} {\it For every right $S_m$-module $N$
 $$ H(N\otimes_{S_m}E_m)=N\otimes_{S_m}H({E_m})$$}
\Pf This result is well-known (it was used implicitly in \cite{Drin1}).
The
proof reduces to the cases of irreducible representations of $S_m$.

Consider an irreducible representation $N_\lambda$ of $S_m$
parametrized by a partition
$\lambda = (\lambda_1^{}, \cdots, \lambda_k^{})$,
$\lambda_1^{} \geq \cdots \geq \lambda_k^{} \ge 0$,
$\sum_{i=1}^k \lambda_i^{} = m$. The symmetric group
$S_m$ acts on the complex $E_m$ and this action is
compatible with the chain map.
So we can split $E_m = \oplus_\lambda E_{m,\lambda}$,
where $E_{m,\lambda}$ is isomorphic to a direct sum of
several (say $m_\lambda$) copies of $N_\lambda^\star$ as
a left $S_m$-module,
where $N_\lambda^\star$ is the contragradient left $S_m$-module
of $N_\lambda$, given by the transpose matrices.
Then, $N_\lambda \otimes_{S_m} E_m \cong
\operatorname{Hom}_{S_m}(N_\lambda^\star, E_{m})$
and so, by Schur's lemma,
$N_\lambda \otimes_{S_m} E_m \cong
N_\lambda \otimes_{S_m} E_{m,\lambda} \cong
E_{m,\lambda}/S_m$.
Since $S_m$ acts on $H(E_m)$ trivially,
$H(E_{m,\lambda}) = 0$  if $N_\lambda$ is not the trivial module.
Hence $H(N_\lambda \otimes_{S_m} E_m) = 0$
if $N_\lambda$ is not the trivial module.
If $N_\lambda$ is the trivial module, we have
  $H(N_\lambda \otimes_{S_m} E_m) = H(E_m)$. \qed

\subsection{Proof of Proposition 3}

Denote the homogeneous part of $\Z^m$-grading $(1,1,\dots,1)$ of
$\cC_m$ by $\Gamma_m$. The symmetric group $S_m$ acts on the right  on
$\Gamma_m$ by permuting the $m$ strings.

\Pro{6:} {\it The chain  complex ($*_m$) is isomorphic to the
 chain complex
$\Gamma_m\otimes_{S_m}E_m$.}

\Pf An element $C$ of $\cC_m$ of $\Z^m$-grading $(1,\dots,1)$ is just
a Chinese character with $m$ external vertices which are numbered from
 1 to $m$. We
map an element $C\otimes(\theta_1,\dots,\theta_k)$ to the element
$D$ of $\cG_k^m$
with the same underlying Chinese character as that of $C$,
 only $\Theta_i(D)$ is the set of external
vertices whose numbers are
 in $\theta_i$. It can be verified at once that this is an isomorphism
 between the two complexes.\qed

\Pro {7:} {\it Suppose $\psi\in\cG_3$ satisfying:
\begin{equation}d\psi=0\tag{$C1'$}\end{equation}
\begin{equation}\psi-\psi^{213}-\psi^{132}=0 \tag {$C2'$}\end{equation}
\begin{equation}\psi=-\psi^{321}\tag{$C3'$}\end{equation} Then
there is a
symmetric element $f\in\cG_2$ such that $df=\psi$.}

($f\in\cG_2$ is symmetric if $f^{21}=f$, by definition.)

\Pf  It suffices to consider the case when $\psi$ is homogeneous.
Since $d\psi=0$, if the $\Z$-grading  $k$ of $\psi$ is
greater than 3 then by the
previous proposition there is $f'\in\cG_2^k$ such that  $df'=\psi$.

If $k= 3$, then the $\Z^3$-grading of $\psi$
must be (1,1,1), i.e $\psi\in\Gamma_3$. Consider $f_1, f_2\in\cG_2^3$
 with the same underlying
Chinese character as $\psi$, only $\Theta_1(f_1)=\Theta_1(\psi)
\cup \Theta_2(\psi)$,
$\Theta_2(f_1)=\Theta_3(\psi)$, $\Theta_1(f_2)=\Theta_1(\psi)$,
 $\Theta_2(f_1)=\Theta_2(\psi)\cup
\Theta_3(\psi)$. Put $f'=(f_1-f_2)/3$. Then using ($C2'$)
 one checks easily that
$df'=\psi$.

In both cases we have $df'=\psi$ for some element $f'\in\cG_2$. Note
that $d(g^{21})= -(dg)^{321}$ for every $g\in\cG_2$. The sum
$f=(f'+\sigma f')/2$ is a symmetric element. Using ($C3'$) we see that
$df=\psi$.\qed

Now Proposition 3 follows from this proposition and Theorem 9.

\section{Invariants of framed oriented links coming from quantum groups}

Suppose for $1\le i,j,k,l\le N$ there are given complex numbers
$r_{ij}^{kl}$. By a state of a chord diagram $D$ in $\cA$  we mean a
map from the set of all arcs of the loop divided by vertices of chords
to the set $\{1,2,\ldots,N\}$. For a fixed state we associate to every
chord of $D$
 a number as indicated below:

\centerline{ \begin{picture}(80,40) \put(0,20){\dashbox{2}(30,0){}}
\thicklines \put(0,35){\vector(0,-1){30}}
\put(30,35){\vector(0,-1){30}} \put(-4,36){\makebox(0,0){$i$}}
\put(35,36){\makebox(0,0){$j$}} \put(35,3){\makebox(0,0){$l$}}
\put(-4,3){\makebox(0,0){$k$}}
\put(60,15){\makebox(0,0){$\Longrightarrow$}}
\put(80,15){\makebox(0,0){$r_{kl}^{ij}$}} \end{picture}}

Take the product of all the numbers associated to all the chords, and
then sum up over all the possible states to get a number. This number
is well-defined (because  of 4-term relation) iff (cf.\cite{Lin,Bar}):

a)$r_{ij}^{kl}=r_{ji}^{lk}$,

b)$[r^{(12)},r^{(13)}]+[r^{(12)},r^{(23)}]=0$.

Where in b) we view $r$ as a linear mapping from $\C^N\otimes\C^N$ to
 $\C^N\otimes\C^N$, and $r^{(ij)}$ is the linear mapping from
 $\C^N\otimes \C^N\otimes\C^N$ to $\C^N\otimes\C^N\otimes\C^N$  which
is as $r$ on the $i$-th and
 $j$-th components while leaves the rest unchanged. The equation b) is
the linearized classical Yang-Baxter equation (\cite{Drin3}).

Suppose $r$ satisfies $a),b)$. Multiplying  $r$ by a formal parameter $h$
and
 applying the above procedure we get for every diagram $D\in\cA$ an
 element $W_r(D)$ in $\C[h]$. If $K$ is a  framed link then
$W_r(\hat{\Zf}(K))\in\C[[h]]$ is an  isotopy invariant.

Now suppose $\g$ is a classical simple Lie algebra, $\rho:\g\rightarrow
 End(V)$ is a representation. Fix an invariant  non-degenerate
 symmetric  bilinear form (Killing form) on $\g$. Let $t$ be the
 symmetric invariant
 element in $\g\otimes \g$ corresponding to the bilinear form.
 Then it can be checked easily that
 $\rho(t)\in End(V)\otimes End(V)$ satisfies both equations $a),b)$.
 Hence we can get an
invariant of framed links $\kappa_{\g,\rho}=W_{\rho(t)}(\hat Z_f)$
 by the above procedure
($t$ is defined up to a constant).

On the other hand, for every representation $\rho:\g\to\ End(V)$,
 using the universal $R$-matrix, one can construct
 another invariant $\tau_{g,\rho}$ of framed links by the
Reshetikhin-Turaev method (cf. \cite{Re-Tu,Tu}). Actually this method
gives a representation of $tangles$ rather than q-tangles and can be
 summarized as follows. There is a structure of {\it ribbon Hopf
algebra}
(\cite{Re-Tu}) on the $h$-adic completion  $\hat U\!\g$ of $U\!
\g\otimes
 \C[[h]]$, where $U\!\g$ is the universal enveloping algebra of $\g$.
 The $R$-matrix of this ribbon  Hopf algebra was constructed by
 Drinfeld and Jimbo \cite{Drin3,Jim} and has expansion $1 + ht/2 +
 O(h^2)$.  The standard procedure
(see \cite{Re-Tu}) associates to  every  representation
$\rho:\g\to  End(V)$   an invariant $\tau_{\g,\rho}$ of
framed oriented links.

\Thm {10:} {\it The two invariants $\kappa_{\g,\rho}$ and
 $\tau_{\g,\rho}$  of framed oriented links are the same,
assuming both are normalized in such a way that both  take
 the value 1 on the empty knot.}

\Pf To see that the two invariants  $\kappa,\tau$ are the same
(problem 4.9 in \cite{Bar}) we can proceed as follows.
 Let $g_1,\dots,g_n$ be an orthonormal base with respect to
the Killing
form. We will first  define a linear mapping $\mu:\cB_m\to
\hat U\!\g^{\otimes m}[[h]]$ .
 Suppose the vertices of a chord diagram $D\in\cB_m$ are
 $a^i_1,\dots,a_{k_i}^i$ on the $i$-th string (the order follows the
 orientation of the string). {\it A state
} is a mapping  $\sigma$ from the set of all vertices $\{a_i^j\}$ to
$\{1,2,\dots,n\}$ which takes the same value on the two vertices of
every chord ($n$ is the dimension of $\g$).  Let $$\mu(D)=h^{(\# \text{
of vertices})/2}\sum_{\text{states $\sigma$}}g_{\sigma(a_1^1)}\dots
g_{\sigma(a^1_{k_1})}\otimes\dots\otimes g_{\sigma(a_1^m)}\dots
g_{\sigma(a^m_{k_m})} $$ This is a well-defined linear mapping
(see also \cite{Bar}).

 Drinfeld proved that (\cite{Drin1,Drin2}) there is another
structure on   $\hat U\!\g$
 which makes   $\hat U\!\g$ a  {\it quasi-triangular quasi-Hopf
 algebra} (not Hopf algebra), with  the
usual co-multiplication of the universal enveloping algebra,
$R=\exp(ht/2)$, $\Phi=\Phi_{KZ}(t^{12},t^{23})$.
Moreover this quasi-triangular quasi-Hopf algebra is a
{\it ribbon quasi-Hopf algebra} (see  the definition of
 ribbon quasi-Hopf algebra in \cite{Coste}), the ribbon
 element is $v=\exp(-\sum_{i=1}^ng_ig_i/2)$.

The series of algebras $\cB_n$ is not a ribbon quasi-Hopf algebra,
 but we have defined operators $\Delta, \ve$, elements $\Phi,R$ for
them. It is easy to see that the mapping $\mu$ commutes with
$\Delta,\ve, \Phi,R$, and the invariant $\kappa_{\g,\rho}$ is
 exactly the invariant of oriented framed links obtained by the
 standard procedure  (see  \cite{Coste}) using the ribbon quasi-Hopf
 algebra and the representation $\rho:\g\to End(V)$.

Drinfeld \cite{Drin1} proved that the above two structures on
 $\hat U\!\g$: 1) ribbon Hopf algebra structure and 2) ribbon
 quasi-Hopf algebra structure  are gauge equivalent, i.e. one can
 be obtained from the other by a  (non-symmetric) twist (see also
\cite{Ko,Kas}). Their categories of representations are equivalent,
 hence the two invariants $\kappa$ and $\tau$ are the same (up to
 constant).\qed

Applying Theorem 10 to
 the trivial knot we get:
$$trace(\exp(h\rho(a)/2))=W_{\rho(t)}(\phi^{-1}) $$
where $a$ is half sum of positive roots.  The left hand side is the value of
$\tau$ on the trivial knot (see \cite{Re-Tu}).

Using the formula in \S4 expressing $\Phi$, and hence $\phi^{-1}$,
in terms of multiple zeta values $\zeta(i_1,\dots,i_k)$, from this
identity we can get many relations  between multiple zeta values.

 The cases $g=sl_N$ or $so_N$ and $\rho$ is the fundamental
representation were treated in detail in \cite{Le-Mu1,Le-Mu2}. In
these
 papers we need not use
 Drinfeld's results to prove Theorem 10, instead we use the
explicit formula of the
Kontsevich integral. Even in these cases the  obtained
identities between multiple zeta values seem far from
 trivial. The famous Euler formula expressing $\zeta(2n)$
in terms of the Bernoulli numbers is among these identities.

\section {Computing the  universal Vassiliev-Kontsevich
 invariant using braids}
It is well known that every framed oriented  link is the closure of a
braid  (figure $a$). For a comprehensive treatment of braids, see
\cite{Bir2}. Let $<\!\beta\!>$ be the framed  oriented link obtained
from a braid $\beta$ by
 closing.  The braid group
 $B_n$ on $n$ strands, with the standard generators
 $\sigma_1,\dots,\sigma_{n-1}$, is the semi-direct product of the pure
 braid group and the symmetric group $S_n$. Denote by $sym$ the
 projection $sym:B_n\to S_n$.

\vskip 6cm

 Regarding every  pure braid as a $q$-tangle, where the top and bottom
 words are defined by the standard order, we get a representation of
 the pure braid group into $\cB_n$ which can be extended to a
 representation of $B_n$ into $\cB_n\rtimes \C[S_n]$. Here
 $\cB_n\rtimes\C[ S_n]$ is the algebra generated by pairs $(D,s), D\in
 B_n, s\in\C[S_n]$ with the  multiplication
 $(D_1,s_1)(D_2,s_2)=(D_1\times s_1(D_2), s_1s_2)$ and bi-linear
 relation (group $S_n$ acts on $\cB_n$ by permuting the strings).  The
 representation is given by
$$\rho(\sigma_1)=(e^{\Omega_{12}/2},sym(\sigma_1)),$$
$$\rho(\sigma_i)=((\Delta^{i-2}\otimes id\otimes id)(\Phi^{-1}\times
e^{\Omega_{23}/2}\times \Phi^{132})\otimes
1^{\otimes(n-i-1)},sym(\sigma_i)) ,$$ for $2\le i\le n-1$ (see also
[Al-co,Drin2,Koh,Piu1]).
 We will compute $\hat Z_f(<\!\beta\!>)$ via $\rho(\beta)$.

Denote by $\cO_n$ the space of chord diagrams on  $n$
 numbered loops. Let $\cO=\oplus_{n=0}^{\infty}\cO_n\otimes \C$. There
is a natural linear mapping from $\cB_n\rtimes \C[S_n]\to \cO$,
$(D,s)\to <\!(D,s)\!>$ by closing  as in figure $b$. For $n=1$ this is
a one-to-one mapping.  Let $\nu\in\cB_1$ be the
 element such that $<\!\nu\!>=\phi^{-1}$. It does not depends on the
 associator $\Phi$ and has rational coefficients.

\Thm {11:}  {\it $\hat{Z}_f(<\!\beta\!>)=<\rho(\beta)(c_n,1)>=
<(c_n,1)\rho(\beta)>$, where $c_n=\Delta^{n-1}(\nu)\in\cB_n$.}

Note that  $c_n$ lies in the center of
$\cB_n$ (Theorem 1) and  $<\!(c_n,1)\!>$ is the chord
diagram without any chord.

Theorem 11 can be proved by evaluating $\hat Z_f$ of the object  $C$
in  figure $c$. It it not a tangle, but one can define $Z_f(C)\in
\cB_n$ in the obvious way. The strings of $C$  are numbered as
indicated in figure $c$.

If one would like to use Theorem 11 as the definition of the universal
Vassiliev-Kontsevich invariant, then one has to check the invariance
under Markov's moves I and II (for the definition of Markov's moves see
[Bir2]). The first move case is trivial due to the fact that $c_n$
belongs to the center of $\cB_n$. The second move case is more
complicated and is equivalent to

\Pro{8:} {\it The following holds true \vskip 8cm}

Here $\Omega=\quad\quad\quad$ and $\omega=\qquad$.  \medskip

This is the exactly the second move case with $n=3$. To get the case of
$n=2$ (or $n>3$) we just apply $\ve\otimes id$  (rep.
$\Delta^{n-3}\otimes id$) to both sides, regarded as elements of
$\cB_2$.

\end{document}